\documentclass[11pt]{article}
\usepackage{Blois,epsfig}

\usepackage{amssymb}
\usepackage{graphicx}
\usepackage{bm}
\usepackage{citesort}
\usepackage{makeidx}
\usepackage{multicol}
\usepackage{bbm}
\usepackage{amsmath}
\usepackage{epsfig}
\long\def\comment#1{ }

\newcommand{\beq}{\begin{eqnarray}}
\newcommand{\eeq}{\end{eqnarray}}
\newcommand{\be}{\begin{eqnarray}}
\newcommand{\ee}{\end{eqnarray}}

\newcommand{\Lam}{\Lambda_{{\rm QCD}}}

\newcommand{\lan}{\langle}
\newcommand{\ran}{\rangle}

\def\simge{\mathrel{%
   \rlap{\raise 0.511ex \hbox{$>$}}{\lower 0.511ex \hbox{$\sim$}}}}
\def\simle{\mathrel{
   \rlap{\raise 0.511ex \hbox{$<$}}{\lower 0.511ex \hbox{$\sim$}}}}
\def\bigs{\mathrel{
   \rlap{\raise 0.531ex \hbox{$>$}}{\lower 0.531ex \hbox{$<$}}}}

\def\empile#1\over#2{\mathrel{\mathop{\kern 0pt#1}\limits_{#2}}}

\begin{document}

\vspace*{4cm}
\title{FROM HIGH--ENERGY QCD TO STATISTICAL PHYSICS}

\author{Edmond IANCU
}

\address{Service de Physique Th\'eorique, CEA/DSM/SPhT,
CE Saclay, F-91191 Gif-sur-Yvette, France}

\maketitle\abstracts{I discuss recent progress in understanding
the high--energy evolution in QCD, which points towards a
remarkable correspondence with the reaction--diffusion problem of
statistical physics. This enables us to determine the asymptotic
behaviour of the scattering amplitudes in QCD.}



\section{Introduction}

Over the last year, an intense activity in the field of
high--energy QCD has been triggered by the following observations:
\texttt{(i)} the gluon number fluctuations in the dilute regime at
low energy play an important role in the evolution towards gluon
saturation and the unitarity limit with increasing energy
\cite{IM032,MS04}, \texttt{(ii)} the QCD evolution in the presence
of fluctuations and saturation is a classical stochastic process
which is similar to the `reaction--diffusion' problem widely
studied in the context of statistical physics \cite{MP03,IMM04},
and \texttt{(iii)} the relevant fluctuations are however missed
\cite{IT04} by the existing approaches to non--linear evolution in
QCD at high energy, namely, the Balitsky--JIMWLK equations
\cite{B,JKLW,RGE,W}. These observations, together with their
consequences, have entailed important conceptual clarifications
and stimulated new ideas and theoretical constructions.

The correspondence between high--energy QCD and statistical
physics was in fact anticipated by the probabilistic structure
inherent in previous approaches like the {\em color dipole
picture}\,\cite{AM94,Salam95,IM031} and the {\em color glass
condensate} \cite{MV,RGE} (the QCD effective theories at low and
high gluon density, respectively). The recent developments in
Refs.\,\cite{MP03,IMM04,IT04} made this correspondence more
precise and also useful (in the sense of generating new results
for QCD), first at the level of the {\em mean field approximation}
--- where the link \cite{MP03} between the Balitsky--Kovchegov
(BK) equation \cite{B,K} in QCD and the
Fisher--Kolmogorov--Petrovsky--Piscounov (FKPP) equation
\cite{FKPP} in statistical physics has shed a new light on the
important phenomenon of geometric scaling \cite{geometric,SCALING}
---, then in the analysis of the {\em particle number
fluctuations} --- where recent advances in statistical physics
\cite{BD,Saar} have enabled us to compute QCD scattering
amplitudes under {\em asymptotic} conditions (very high energy and
arbitrarily small coupling constant) \cite{IMM04,IT04}. These new
approaches have elegantly confirmed and extended previous results
obtained through direct studies in QCD
\cite{SCALING,MT02,DT02,MS04}.

At the same time, it became clear that the correspondence with
statistical physics cannot be used to also study the {\em
pre}--asymptotic behaviour in QCD, that is, to compute scattering
amplitudes for realistic values of the energy and the coupling
constant. In that regime, which is the only one to be interesting
for the phenomenology, one rather needs the actual evolution
equations in QCD at high energy. As aforementioned, these
equations should be more general --- in the sense of also
including the effects of gluon number fluctuations --- than the
previously known Balitsky--JIMWLK equations. So far, the relevant
equations have been constructed\,\cite{IT04,MSW05,BIIT05} only in
the limit where the number of colors $N_c$ is large. An ambitious
program, which aims at generalizing these equations to arbitrary
values of $N_c$, is currently under way
\cite{KL05,BREM,MMSW05,HIMS05,Balit05}. This effort led already to
some important results  --- in particular, the
recognition\,\cite{KL05,BIIT05} of a powerful `self--duality'
property of the high--energy evolution, and the construction of an
effective Hamiltonian which is explicitly self--dual
\cite{BREM,Balit05}
---, but the general problem is still under study, and the
evolution equations for arbitrary $N_c$ are not yet known.

In my two succinct contributions to these Proceedings, I shall
restrict myself to the large--$N_c$ limit, which is quite
intuitive in that it allows the use of a suggestive dipole
language \cite{AM94,IT04}. In this context, I shall rely on simple
physical considerations to explain the correspondence between
high--energy QCD and statistical physics (in this presentation),
and then motivate the structure of the recently derived `evolution
equations with Pomeron loops' (in my other presentation
\cite{Blois2}).

\section{QCD evolution at high energy}
\begin{figure}
\begin{center}
\centerline{\epsfig{file=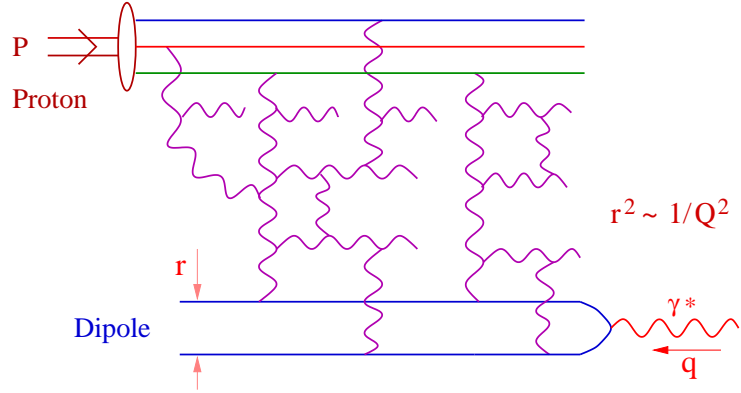,height=5.2cm}}
\caption{\sl An instantaneous gluon configuration in the proton
wavefunction as `seen' in DIS at small $x$.
    \label{dis_blois1}}
\end{center}
\end{figure}

To put the theoretical developments into a specific physical
context, let us consider $\gamma^*$--proton deep inelastic
scattering (DIS) at high energy, or small Bjorken--$x$. We shall
view this process in a special frame in which most of the total
energy is carried by the proton, whose wavefunction is therefore
highly evolved, while the virtual photon has just enough energy to
dissociate long before splitting into a quark--antiquark pair in a
colorless state (a `color dipole'), which then scatters off the
gluon distribution in the proton (see Fig. \ref{dis_blois1}). The
transverse size $r$ of the dipole is controlled by the virtuality
$Q^2$ of $\gamma^*$ (roughly, $r^2\sim 1/Q^2$), so for $Q^2\gg
\Lam^2$ one can treat the dipole scattering in perturbation
theory. But for sufficiently small $x$, even such a small dipole
can see a high--density gluonic system, and thus undergo strong
scattering.

Specifically, the small--$x$ gluons to which couple the projectile
form a {\em color glass condensate\,\,}\cite{MV,RGE} (CGC), i.e.,
a multigluonic state which is characterized by high quantum
occupancy, of order $1/\alpha_s$, for transverse momenta $k_\perp$
below the {\em saturation momentum} $Q_s(x)$, but which becomes
rapidly dilute when increasing $k_\perp$ above $Q_s$. The
saturation scale rises very fast with the energy\,\cite{DT02},
$Q_s^2(x)\sim x^{-\lambda}$, and is the fundamental scale in QCD
at high energy. In particular, a small external dipole with size
$r\ll 1/Q_s$ undergoes only weak scattering (since it couples to
the dilute tail of the gluon distribution at large  $k_\perp$),
while a relatively large dipole with $r\simge 1/Q_s$ `sees' the
saturated gluons, and thus is strongly absorbed.

\begin{figure}[t]
    \centerline{\hspace{1.cm}\epsfxsize=3.7cm\epsfbox{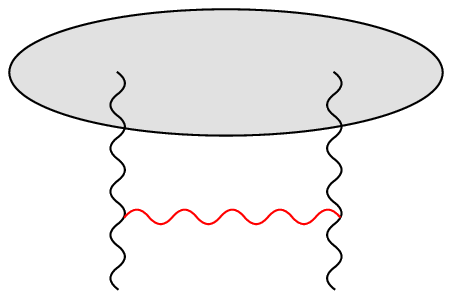}
    \hspace{.3cm}\epsfxsize=4.cm\epsfbox{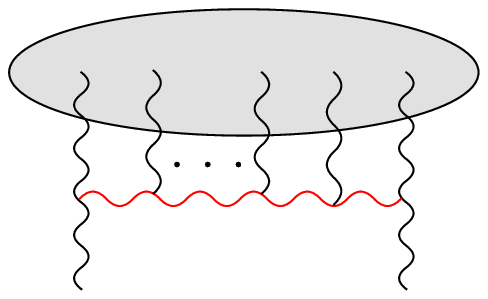}
    \hspace{.3cm}\epsfxsize=4.cm\epsfbox{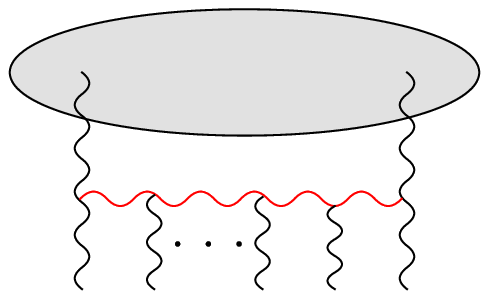}}
 \vspace*{0.2cm}
 \hspace{3.7cm} (a)\hspace{3.7cm} (b)\hspace{4.cm}(c)
    \caption{\sl Gluon processes which occur
    in one step of high energy
    evolution. \label{BREMfig}}
\end{figure}

In turn, the small--$x$ gluons are produced through `quantum
evolution', i.e., through radiation from {\em color sources}
(typically, other gluons) at larger values of $x$, whose internal
dynamics is `frozen' by Lorentz time dilation. Let $\tau =\ln 1/x$
denote the {\em rapidity\,}; it takes, roughly, a rapidity
interval $\Delta\tau \sim 1/\alpha_s$ to emit one small--$x$
gluon; thus, in the high energy regime where $\alpha_s\tau \gg 1$,
the dipole meets with well developed gluon cascades, as shown in
Fig. \ref{dis_blois1}. Three types of processes can be
distinguished in Fig. \ref{dis_blois1}, which for more clarity are
disentangled in Fig. \ref{BREMfig}.

The first process, Fig. \ref{BREMfig}.a, represents one step in
the standard BFKL evolution \cite{BFKL}; by iterating this step,
one generates gluon ladders which are resummed in the solution to
the BFKL equation \cite{BFKL}. By itself, this mechanism entails a
rapid growth of the gluon distribution (exponential in $\tau$),
which however leads to conceptual difficulties at very high energy
: \texttt{(i)} The BFKL estimate for the dipole scattering
amplitude $T_\tau(r)$ grows like a power of the energy, and thus
eventually violates the unitarity bound $T_\tau(r)\le 1$. (The
upper limit $T_\tau=1$ corresponds to the `black disk' limit, in
which the dipole is totally absorbed by the target.) \texttt{(ii)}
The BFKL ladder is not protected from deviations towards the
non--perturbative domain at low transverse momenta
$k_\perp^2\simle \Lam^2$ (`infrared diffusion'). With increasing
energy, the BFKL solution receives larger and larger contributions
from such soft intermediate gluons, and thus becomes unreliable.

These `small--$x$ problems' of the BFKL evolution are both cured
by {\em gluon saturation} \cite{GLR}, the mechanism leading to the
formation of a CGC: At sufficiently high energy, when the gluon
density in the target becomes very large, the $n\to 2$ {\em
recombination} processes illustrated in Fig. \ref{BREMfig}.b start
to be important and tame the growth of the gluon distribution.
Such processes are included (to all orders) in the JIMWLK equation
\cite{JKLW,RGE,W}, a non--linear and  functional generalization of
the BFKL equation which describes the evolution of the ensemble of
gluon correlations in the approach towards saturation. (As
manifest on Fig. \ref{BREMfig}.b, the standard `gluon
distribution', which is a 2--point function, gets coupled to the
higher $n$--point functions via the recombination effects.)
Remarkably, the JIMWLK equation is a (functional) {\em
Fokker--Planck equation}, which describes the high--energy
evolution as a {\em classical stochastic process} \cite{PATH} ---
a random walk in the functional space of gluon fields. When
applied to scattering amplitudes for simple external projectiles
(like color dipoles, quadrupoles, etc.), the JIMWLK equation
generates an infinite set of coupled evolution equations that were
originally derived by Balitsky \cite{B}.

However, as recently noticed in Ref. \cite{IT04}, the
Balitsky--JIMWLK hierarchy misses the $2\to n$ {\em splitting}
processes illustrated in Fig. \ref{BREMfig}.c, which describe the
{\em bremsstrahlung} of additional small--$x$ gluons in one step
of the evolution. By themselves, such processes are important in
the {\em dilute} regime at relatively low energy (or, for a given
energy, at relatively high transverse momenta $k_\perp\gg Q_s$),
where they generate the $n$--point correlation functions with
$n>2$ from the dominant 2--point function (the gluon
distribution). But once generated, the higher $n$--point functions
are rapidly amplified by their subsequent BFKL evolution (the
faster the larger is $n$) and eventually play an important role in
the non--linear dynamics leading to saturation. Thus, such
splitting processes {\em are} in fact important for the evolution
towards high gluon density, as first observed in numerical
simulations \cite{Salam95} of Mueller's `dipole picture'
\cite{AM94} and recently explained in Refs.
\cite{MS04,IMM04,IT04}.

\section{QCD scattering amplitudes from statistical physics}

Equations including both merging and splitting in the limit where
the number of colors $N_c$ is large have recently became available
\cite{IT04}, but these are still quite complicated and their
exploration is only at the beginning (see my next presentation
\cite{Blois2}). Still, as we shall argue now, the {\em asymptotic}
behaviour of the corresponding solutions --- where by `asymptotic'
we mean both the {\em high--energy} limit $\tau\to\infty$ and the
{\em weak coupling} limit $\alpha_s\to 0$
--- can be {\em a priori} deduced from universality considerations
relating high--energy QCD to problems in statistical physics
\cite{IMM04}.

To that aim, it is convenient to rely on the event--by--event
description \cite{IMM04} of the scattering between the external
dipole and the hadronic target, and to use the large--$N_c$
approximation to replace the gluons in the target wavefunction by
color dipoles  \cite{AM94} (which is indeed correct in the dilute
regime). Then, the scattering amplitude in a given event can be
estimated as
 \be\label{Tf}
  T_{\tau}(r,b)
  \,\simeq \,\alpha_s^2 \,f_{\tau} (r,b)\,,\ee
where $\alpha_s^2$ is the scattering amplitude between two dipoles
with comparable sizes and nearby impact parameters, and $f_{\tau}
(r,b)$ is the {\em occupation number} for target dipoles with size
$r$ at impact parameter $b$. Since, clearly, $f$ is a {\em
discrete} quantity: $f=0,1,2,\dots$, so is also the scattering
amplitude in a given event: $T$ is a multiple integer of
$\alpha_s^2$. The estimate (\ref{Tf}) is based on the single
scattering approximation, and thus is valid in the dilute target
regime, where $ T \ll 1$.

In this dipole language, the $2\to 4$ gluon splitting depicted in
Fig. \ref{BREMfig}.c is tantamount to $1\to 2$ dipole splitting,
and generates {\em fluctuations} in the dipole occupation number
and hence in the scattering amplitude. Thus, the evolution of the
amplitude $T_{\tau}(r,b)$ with increasing $\tau$ represents a {\em
stochastic process} characterized by an expectation value $\lan
T(r,b)\ran_{\tau} \simeq \alpha_s^2 \,\lan f (r,b) \ran_{\tau}$,
and also by fluctuations $\delta T \sim \alpha_s^2\delta f \sim
\sqrt{\alpha_s^2 T}$, where we have used the fact that $\delta f
\sim \sqrt{f}$ for fluctuations in the particle number. These
fluctuations are relatively important (in the sense that $\delta T
\simge  T$) only in the {\em very} dilute regime where $\lan f
\ran\simle 1$, or $\lan T\ran\simle \alpha_s^2$.

\subsection{The mean field approximation (BK, FKPP
\& geometric scaling)}

Unitarity corrections in the form of {\em multiple scattering}
start to be important when $T\sim 1$; according to Eq.~(\ref{Tf}),
this happens for dipole occupation numbers of order
$1/\alpha_s^2$. Consider first the formal limit $\alpha_s^2\to 0$,
in which the maximal occupation number $N\sim 1/\alpha_s^2$
becomes arbitrarily large. Then one can neglect the particle
number fluctuations and follow the evolution of the scattering
amplitude in the {\em mean field approximation} (MFA). This is
described by the BK equation \cite{B,K}, a non--linear version of
the BFKL equation which, mediating some approximations, can be
shown \cite{MP03} to be equivalent to the FKPP equation
\cite{FKPP}. The latter represents the MFA for the
reaction--diffusion process $A \rightleftharpoons A+A$ and related
phenomena in biology, chemistry, astrophysics, etc. (see
\cite{Saar} for recent reviews and more references), and reads
 \be\label{BK}
 \partial_\tau T(\rho,\tau)\,=\,
 \partial_\rho^2 T(\rho,\tau)\,+\,
 T(\rho,\tau)\,- \,T^2(\rho,\tau),\ee
in notations appropriate for the QCD problem at hand:
$T(\rho,\tau)\equiv \lan T(r)\ran_{\tau}$ and $\rho\equiv
\ln(r_0^2/r^2)$, with $r_0$ a scale introduced by the initial
conditions at low energy. Note that weak scattering ($T\ll 1$)
corresponds to small dipole sizes ($r\ll 1/Q_s$), and thus to
large values of $\rho$. In momentum space, $\rho\sim \ln
k_\perp^2$. The three terms on the r.h.s. of Eq.~(\ref{BK})
describe, respectively, diffusion, growth and recombination.
Together, the first two terms represent an approximate version of
the BFKL dynamics, while the latter is the non--linear term which
describes multiple scattering and ensures that the evolution is
consistent with the unitarity bound $T\le 1$. In fact, $T=1$ is
clearly the high--energy limit of the solution to Eq.~(\ref{BK}).

\begin{figure}[t]    
    \centerline{\epsfxsize=15.cm\epsfbox{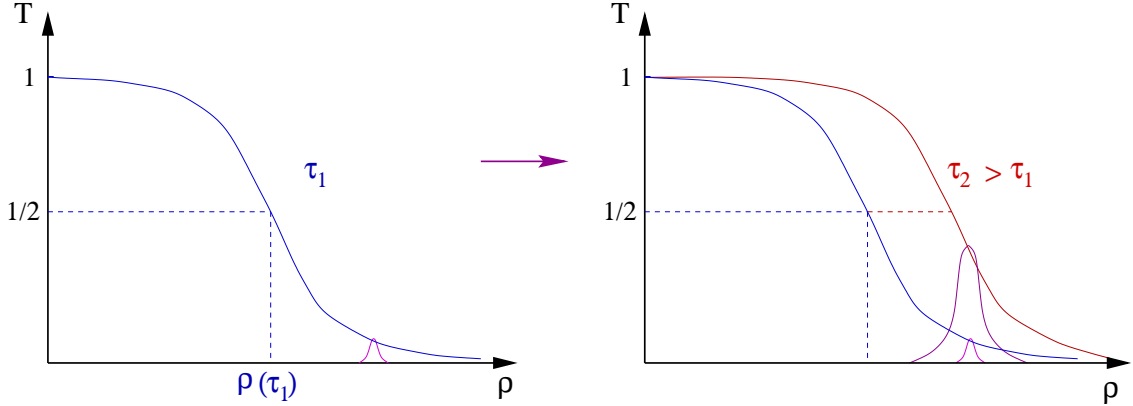}}
    \caption{\sl Evolution of the continuum front of the Balitsky--Kovchegov
    equation with increasing $\tau$.
                 \label{TWave5}}
\end{figure}

The solution $T_\tau(\rho)$ to Eq.~(\ref{BK}) is a {\em front}
which interpolates between two fixed points : the stable fixed
point $T=1$ (the unitarity limit) at $\rho\to -\infty$, and the
unstable fixed point $T=0$ at $\rho\to \infty$ (see Fig.
\ref{TWave5}). The position of the front, which marks the
transition between strong and weak scattering, defines the {\em
saturation scale\,}: $\rho_s(\tau)\equiv \ln(r_0^2 Q_s^2(\tau))$.
With increasing $\tau$, the front moves towards larger values of
$\rho$, as illustrated in Fig. \ref{TWave5}.

The dominant mechanism for front propagation is the BFKL growth in
the tail of the distribution at large $\rho$ : the front is {\em
pulled} by the rapid growth of a small perturbation around the
unstable state\,\cite{Saar}. In view of that, the {\em velocity}
of the front $\lambda\equiv {d\rho_s}/{d\tau}$ is fully determined
by the {\em linearized} version of Eq.~(\ref{BK}) (i.e., the BFKL
equation), which describes the dynamics in the tail. By solving
the BFKL equation one finds\,\cite{SCALING,MT02} that, for $\rho
>\rho_s(\tau)$ and sufficiently large $\tau$,
 \be \label{TBFKL}
 T_\tau(\rho) \,\simeq\,{\rm e}^{\omega \bar\alpha_s \tau}
 \,{\rm e}^{-\gamma\rho}  \,=\,{\rm e}^{-\gamma(\rho -\rho_s(\tau))}\,
 ,\qquad \rho_s(\tau)\equiv c \bar\alpha_s \tau,\ee
where $\bar{\alpha}_s = {\alpha}_s N_c/\pi$, $\gamma=0.63..$, and
$c \equiv \omega/\gamma=4.88..\,$. From Eq.~(\ref{TBFKL}) one can
immediately identify the velocity of the front in the MFA as
$\lambda_0 = c\bar\alpha_s$. Since $Q_s^2(\tau) \simeq Q_0^2\,
{\rm e}^{\lambda_0 \tau}$, it is furthermore clear that
$\lambda_0$ plays also the role of the {\em saturation exponent}
(here, in the MFA).

According to Eq.~(\ref{TBFKL}), the scattering amplitude depends
only upon the difference $\rho-\rho_s(\tau)$ : $T_\tau(\rho)={\cal
A} \big(\rho-\rho_s(\tau)\big)$. This is an exact property of the
FKPP equation (at sufficiently large $\tau$) and expresses the
fact that the corresponding front is a {\em traveling wave} which
propagates without distortion \cite{Saar}. In QCD, this property
is valid only within a limited range, namely for \cite{SCALING}
 \be
  0\, <\,\rho-\rho_s(\tau) \, \simle\, \rho_s(\tau)\,\ee
(the ``geometric scaling window''; see below), because of the more
complicated non---locality of the BFKL, or BK, equations. When
translated to the original variables $r$ and $\tau$, this property
implies that the dipole amplitude {\em scales} as a function of
$r^2Q_s^2(\tau)$ : $\lan T(r)\ran_{\tau}\approx {\cal
A}\big(r^2Q_s^2(\tau)\big)$. This is the property originally
referred to as {\em geometric scaling}
\cite{geometric,SCALING,MT02}, and which might explain a
remarkable regularity observed \cite{geometric} in the small--$x$
data for DIS at HERA. Namely, for $x\le 0.01$, the total
cross--section $\sigma_{\gamma^*p}(x,Q^2)$ for the absorbtion of
the virtual photon shows approximate scaling as a function of
$Q^2/Q_s^2(x)$, with $Q_s^2(x) \propto (1/x)^\lambda$ and $\lambda
\approx 0.3$ from a fit to the data. This measured value of
$\lambda$ is quite far away from the above prediction $\lambda_0 =
c\bar\alpha_s\sim 1$ of the BFKL equation; but after including the
NLO corrections to the BFKL equation (see Ref. \cite{DT02} for
details), the ensuing, improved, theoretical prediction
\cite{DT02} decreases indeed to a value close to 0.3.

\subsection{The effects of fluctuations}

\begin{figure}[t]    
    \centerline{\epsfxsize=14.cm\epsfbox{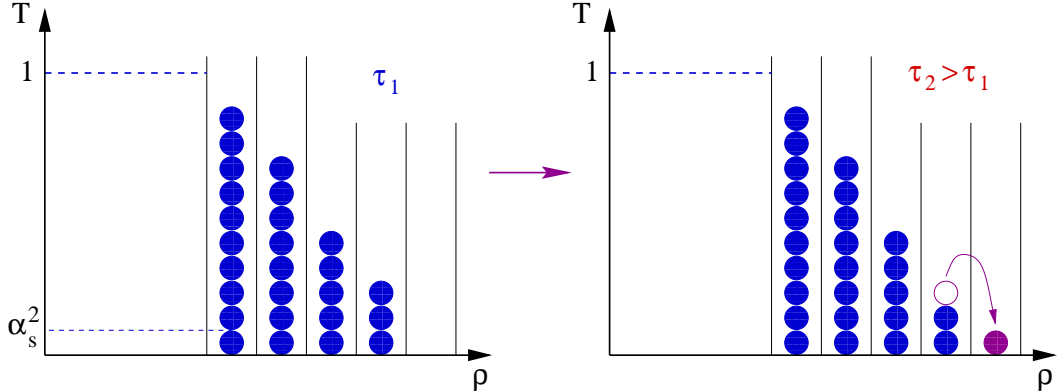}}
    \caption{\sl Evolution of the discrete front of a microscopic
    event with increasing rapidity $\tau$. The small blobs
    are meant to represent the elementary
    quanta $\alpha_s^2$ of $T$ in a microscopic event.
                 \label{TWave6}}
\end{figure}

What is the validity of the mean field approximation ? We have
earlier argued that the gluon splitting processes (cf. Fig.
\ref{BREMfig}.c) responsible for dipole number fluctuations should
play an important role in the dilute regime. This is further
supported by the above considerations on the {\em pulled} nature
of the front: Since the propagation of the front is driven by the
dynamics in its tail where the fluctuations are {\em a priori}
important, the front properties should be strongly sensitive to
fluctuations. This is indeed known to be the case for the
corresponding problem in statistical physics \cite{BD,Saar}, as it
can be understood from the following, qualitative, argument:

Consider a particular realization of the stochastic evolution of
the target, and the corresponding scattering amplitude, which is
discrete (in steps of $\alpha_s^2$). Because of discreteness, the
microscopic front looks like a histogram and thus is necessarily
{\em compact} : for any $\tau$, there is only a finite number of
bins in $\rho$ ahead of $\rho_s$ where $T_\tau$ is non--zero (see
Fig. \ref{TWave6}). This property has important consequences for
the propagation of the front. In the empty bins on the right of
the tip of the front, the local, BFKL, growth is not possible
anymore (as this would require a seed). Thus, the only way for the
front to progress there is via {\it diffusion}, i.e., via
radiation from the occupied bins at $\rho <\rho_{\rm tip}$
(compare in that respect Figs. \ref{TWave5} and \ref{TWave6}). But
since diffusion is less effective than the local growth, we expect
the velocity of the microscopic front (i.e., the saturation
exponent) to be reduced as compared to the respective prediction
of the MFA.

To obtain an estimate for this effect \cite{MS04,IMM04}, we shall
rely on the universality of the dominant {\em asymptotic}
($\tau\to\infty$ and $N\equiv 1/\alpha_s^2\gg 1$) behaviour, which
has been observed in the context of statistical physics and
justified by Brunet and Derrida \cite{BD} through the following,
intuitive, argument: For a given microscopic front and $N\gg 1$,
the MFA should work reasonably well everywhere except in the
vicinity of the tip of the front, where the occupation number $f$
becomes of order one and the linear growth term becomes
ineffective. (Note that, in QCD, $f\sim 1$ corresponds to $T\sim
\alpha_s^2$, which is precisely where one expects the fluctuation
effects to become important.) Accordingly, Brunet and Derrida
suggested a modified version of the FKPP equation (\ref{BK}) in
which the `BFKL--like' growth term is switched off when $T<
\alpha_s^2$ :
 \be\label{BKBD}
 \partial_\tau T(\rho,\tau)\,=\,
 \partial_\rho^2 T\,+\,\Theta\big(T - \alpha_s^2\big)
 T(1- T).\ee
By solving this equation in the linear regime, they have obtained
the first correction to the front velocity as compared to the MFA
(in notations adapted to QCD; see Ref. \cite{IMM04} for details):
 \be\label{ls}
 \lambda\,\simeq\,\bar\alpha_s\left[c\,-\,
 \frac{\kappa}{\ln^2(1/\alpha_s^2)}\,+\,{\cal O}
 \big(1/\ln^3 \alpha_s^2\big)\right]
 \,,
 \ee
where the numbers $c \approx 4.88$ and $\kappa \approx 150$ are
fully determined by the linear (BFKL) equation. In QCD, the same
result has been first obtained through a different but related
argument by Mueller and Shoshi \cite{MS04}. Note the extremely
slow convergence of this result towards its mean field limit: the
corrective term vanishes only logarithmically with decreasing
$1/N=\alpha_s^2$, rather than the power--like suppression usually
found for the effects of fluctuations. This is related to the high
sensitivity of the pulled fronts to fluctuations, as alluded to
above. The merely logarithmic dependence of Eq.~(\ref{ls}) upon
the value of the cut--off also explains its universality: a
renormalization $\alpha_s^2 \to A\alpha_s^2$ of the latter does
not change the dominant correction in Eq.~(\ref{ls}).

But although it becomes an {\em exact} result in QCD in the formal
limit $\alpha_s^2 \to 0$, the estimate in Eq.~(\ref{ls}) is
clearly useless for any practical application, because of the very
slow convergence of the expansion there. Still, this has the merit
to demonstrate that the effects of the fluctuations are
potentially large, which invites us to critically reexamine the
results previously obtained from the BK equation. In fact, from
the correspondence with statistical physics, we also know
\cite{IMM04} that the geometric scaling property of the BK
solution will be eventually washed out by fluctuations at
sufficiently high energy. But in order to understand how fast this
actually happens (i.e., up to what energy one should expect
geometric scaling to be a good property), and also to estimate the
saturation exponent for realistic values of $\alpha_s$, one needs
to solve the actual evolution equations in QCD at large--$N_c$
\cite{IT04,MSW05}, to be described in my next contribution to
these Proceedings \cite{Blois2}.


\end{document}